# Reionization and Cosmic Microwave Anisotropies

Naoshi Sugiyama[1,3], Joseph Silk[1] and Nicola Vittorio[2]

[1] *Departments of Astronomy and Physics, and Center for Particle Astrophysics, University of California, Berkeley, CA 94720*

[2] *Dipartimento di Fisica, Universita di Roma 'Tor Vergata', Viale della Ricerca Scientifica, 00173 Roma, Italy*

[3] *Department of Physics, Faculty of Science, University of Tokyo, Tokyo 113, Japan*

## ABSTRACT

The effects of reionization, occurring after standard recombination in cold dark matter-dominated models, on CMB anisotropies are investigated. Late-time reionization reduces the CMB anisotropies, in particular, on degree scales. It is found that constraints on cold dark matter-dominated models from the highest frequency channel of the 9-point South Pole data are significantly relaxed for models which are consistent with Big Bang nucleosynthesis if reionization is assumed to have occurred by redshift $\sim 20$.

*Subject headings*: cosmology: cosmic microwave background - dark matter

e-mail: sugiyama@bkyast.berkeley.edu
silk@pac2.berkeley.edu
vittorio@roma2.infn.it
*Astrophysical Journal Letter* to be published.

# I. Introduction

The discovery of cosmic microwave background (CMB) anisotropies by the *COBE* satellite (Smoot et al. 1992) has provided important information about initial conditions in terms of the spectrum of primordial density perturbations. In order to know the consistency of specific cosmological models that are relevant for large scale structure formation, however, we require observations of CMB anisotropies on scales smaller than 10 degrees. On these scales, the effects of possible reionization on CMB anisotropies after the standard epoch of hydrogen recombination are expected to be important.

There are several reasons for considering late-time reionization of the whole universe. From the Gunn-Peterson test, we know the universe must have been highly reionized before $z = 5$. Studies of the intracluster gas abundances imply that there was considerable injection of metals, and hence energy, into the intergalactic medium during the earliest stages of galaxy formation. Last but not least, the recent claims that cold dark matter-dominated models (CDM models) are unacceptable (Gorski, Stompor and Juszkiewicz 1993; Muciaccia et al., 1993) or severely constrained (Dodelson and Jubas 1993), as a consequence of the recent observation at the South Pole by Gaier et al. (1992) (hereafter as SP91) means that it is especially timely to carefully investigate the extent to which reionization reduces CMB anisotropies on small and intermediate scales. The beam of this latter experiment had a sinusoidal throw on the sky with amplitude $1.°5$. This experiment is particularly constraining for CDM models because the CMB power spectrum in such models has a high peak around the one degree scale which corresponds to the horizon scale of the last scattering surface. However if early reionization occurred, this high peak is smeared out. In section 2, we will discuss this point in detail for CDM models.

The possibility that CMB anisotropies were smeared out by reionization of the early universe was pointed out by Vittorio and Silk (1984) and Bond and Efstathiou (1984). However, these pioneering papers only estimated the angu-lar scale of such smearing. Bond et al. (1989) numerically calculated CMB anisotropies in a reionized universe. They investigated models with no recombination. However such models are incompatible with COBE-normalized CDM: reionization cannot plausibly occur before $z \sim 100$ with any reasonable efficiency for early star formation and ionizing photon production (Tegmark, Silk and Blanchard 1993). Moreover models with no recombination come close to violating (Tegmark and Silk 1994) the recent constraint on the Compton $y$ parameter $y < 2.5 \times 10^{-5}$ set by the *COBE* FIRAS detector (Mather et al. 1993). Hence we consider here more realistic reionization at relatively late epochs, and investigate the behaviour of CMB anisotropies in such models. We also compare CMB anisotropies with the SP91 experiment by using *COBE* observations to normalize the perturbation amplitude, and derive constraints on CDM models imposed by reionization.

In section 2, we will show how reionization after recombination affects the CMB anisotropies. Constraints on CDM models are given from SP91 data in section 3. Conclusions and discussions are presented in section 4.

## II. Behaviour of CMB Anisotropies in a Reionized Universe

Throughout this paper, we only consider *standard* models, i.e., CDM models with density parameter $\Omega = 1$, Harrison-Zel'dovich initial spectrum, initially adiabatic perturbations, and baryon density within the range allowable by primordial nucleosynthesis. The initial conditions are set well before the epoch of equal densities of matter and radiation, that demarcates the onset of subhorizon fluctuation growth. We use the gauge invariant method (Bardeen 1980, Kodama and Sasaki 1984) to treat perturbation variable. The perturbation equations are solved numerically up till the present epoch. The photon distribution function is calculated by using the multipole expansion method (see e.g., Sugiyama and Gouda 1992). The temperature anisotropy is expanded in Fourier space by

writing

$$\frac{\delta T}{T}(\mathbf{k},\boldsymbol{\gamma}) = \sum_{\ell \geq 2}(-i)^{\ell}(2\ell+1)a_{\ell}(k)P_{\ell}(\cos\theta),$$

where $\mathbf{k}$ and $\boldsymbol{\gamma}$ are the wave number vector and direction vector of the photon propagation. Here, $k$ is the amplitude of the wave number and the direction cosine is defined by $\cos\theta \equiv \boldsymbol{\gamma} \cdot \mathbf{k}/k$. The rms temperature anisotropies are defined by

$$< \left(\frac{\delta T}{T}\right)^2 >_{rms} = \frac{1}{2\pi^2}\int dk k^2 \sum_{\ell \geq 2}(2\ell+1)|a_{\ell}(k)|^2.$$

The features of the CMB spectrum in standard CDM models without reionization are as follows. Denote the density fluctuation amplitude by $\delta\rho/\rho$: the corresponding potential and velocity perturbations are given in the Newtonian limit by solving the Poisson and continuity equations,

$$\nabla^2 \delta\phi = 4\pi G \,\delta\rho; \quad \nabla \cdot \delta\mathbf{v} = -\frac{\partial}{\partial t}\left(\frac{\delta\rho}{\rho}\right).$$

The spatial derivatives are evaluated over subhorizon scales and are in comoving coordinates. On large scales, the Sachs-Wolfe effect (Sachs and Wolfe 1967) dominates: $\delta T/T = \delta\phi/3c^2$. We can see the plateau of the Sachs-Wolfe effect on the CMB spectrum in Figure 1. Here the cutoff on very large scales is because of subtraction of the monopole and dipole components. There is a high peak, the so-called Doppler peak: $\delta T/T = (\delta v/c)_{LS}$ at intermediate scales on the last scattering surface (Sunyaev and Zeld'vich 1970). This peak corresponds to the horizon scale at the last scattering surface. If reionization did not occur or occurred at a very recent epoch, scattering between photons and electrons is not effective after the recombination epoch and the universe is transparent. Hence the scale of the last scattering surface is the horizon scale at the recombination epoch. On small scales, the adiabatic term dominates: $\delta T/T \approx (1/3)(\delta\rho/\rho)_{LS}$, and there is significant damping because of photon diffusion (Silk 1968). This scale corresponds to the thickness of the last scattering surface.

If reionization occurs early enough, however, the spectrum shows a different feature. In Figure 2, the time evolution is shown of perturbations in the reionized universe. The scale of perturbations coincides with the Doppler peak in Fig.1. It is found that the amplitude of CMB anisotropies grows after entering the Jeans scale. Once recombination occurs, however, this perturbation stops growing and remains constant because of the free streaming behaviour. Then, if we assume reionization at $z = 100$, the perturbation is suddenly damped as shown in this figure. As a result, CMB anisotropies on scale smaller than the horizon scale of the new last scattering surface are smeared out. This new last scattering surface occurs where the optical depth equals unity. For CDM models, this occurs in the range $z = 50 - 100$. The resultant spectrum is also shown in Fig.1. The original Doppler peak is strongly suppressed. However, we have found an unexpected result: the Doppler peak is recreated on a larger scale, corresponding to the horizon scale at the new last scattering surface. There are also contributions to the temperature fluctuations on small angular scales due to second order perturbations via the Vishniac effect (Ostriker and Vishniac 1986; Vishniac 1987). However this effect is less important for CDM models than for isocurvature baryon fluctuation models (Hu, Scott and Silk 1993).

In order to directly compare the spectrum with specific observations, we show the coefficients $C_\ell$ of the CMB anisotropies in $\ell$ space as a parameter of the epoch of reionization for a CDM model in Figure 3a. $C_\ell$ is defined by

$$C_\ell \equiv \frac{2}{\pi} \int dk k^2 |a_\ell|^2 .$$

The expected temperature anisotropy for each experiment is expressed by using $C_\ell$ and the specific window function $W_\ell$ (Bond et al. 1989):

$$\left(\frac{\delta T}{T}\right)^2_{exp} = \sum_{\ell \geq 2} \frac{2\ell + 1}{4\pi} C_\ell W_\ell .$$

The anisotropies are normalized to the COBE $10°$ observation, which we take to be $(\delta T/T)_{rms} = 30 \mu K$. The contribution of the Vishniac effect is also shown in this figure where we assume the universe is optically thick. In Figure 3b, the window functions of SP91, OVRO (Readhead et al. 1989) and COBE are shown.

Evidently, reionization is effective at significantly reducing the Doppler peak in Figure 3a. The height of the peak is less than 70% of that in the standard (no reionization) model even for reionization as late as $z = 20$. There seems to be little difficulty with COBE-normalized CDM , for reionization of the universe to substantially reduce the expected temperature anisotropies on scales smaller than a few degrees.

## III. Constraints on Models

Our reionization calculations may be used to constrain the baryon density and Hubble constant parameter space if we adopt the nucleosynthesis prediction that $\Omega_b h^2 = 0.02 \pm 0.01$. This range is wider than quoted by Walker et al. (1991), to allow for possible sources of systematic uncertainty (*e.g.* Schramm 1993). We ask the question: what reionization redshift is necessary to bring the SP91 constraint into agreement with CDM?

We assume that the universe becomes suddenly fully ionized at some epoch after the recombination. We allow the baryon density parameter $\Omega_B$, the nondimensional Hubble constant $h$ normalized by 100km/s/Mpc, and the epoch of reionization $z_{re}$ to be free parameters, for a specified CMB fluctuation limit.

In order to obtain the most severe constraints, we concentrate on the SP91 experiment because this experiment is the most stringent for CDM models (Gorski, Stompor and Juszkiewicz 1993; Muciaccia et al., 1993; Dodelson and Jubas 1993). We use the observed value $30\mu K$ at $10°$ by COBE to normalize the CMB anisotropies. In order to treat SP91, we use the Bayesian method (Bond et al. 1989) and assume a uniform prior, although this assumption has been criticized by Bunn et al. (1993) as introducing additional uncertainty. We take into account the beam profile and subtraction of the mean and the gradient of the temperature anisotropies. Only the highest frequency channel data are considered because the data in the other channels seem to be contaminated by foreground galactic emission (Gaier et al. 1992). Note that this also introduces further uncertainty because of the small number of independent data points and the relatively narrow range of frequencies in the SP91 experiment. We also verified that the expected values of $\sigma_8$ and the quadrupole anisotropy do not depend on the thermal history of the universe after recombination because the CDM fluctuation spectrum is unaffected by the thermal history, and the quadrupole anisotropy is far larger than the horizon scale at last scattering. These values of $\sigma_8$ and the quadrupole anisotropy are consistent with the observed range.

Ignoring any additional uncertainties, the Bayesian constraints from the high frequency channel of SP91 on the $\Omega_B - h$ plane are shown in Figure 4. If reionization occurred at $z = 10$, the constraint is almost the same as for the models without reionization (Dodelson and Jubas 1993). However, even reionization at $z = 20$ significantly weakens the constraint on $\Omega_B$. All models which are consistent with Big Bang nucleosynthesis survive, even though the universe is far from being optically thick at last scattering.

## IV. Conclusions

We have investigated effects of late time reionization on CMB anisotropies. Such reionization smooths out the original temperature fluctuations on scales smaller than a few degrees while it creates new fluctuations on larger scales that correspond to the horizon scale at last scattering, as well as on much smaller scales by the Vishniac effect. We have compared the expected anisotropies for standard CDM models with SP91. The reionization epoch was varied as a free parameter. In order to save large $\Omega_B$ models which are consistent with Big Bang nucleosynthesis, we found that reionization only as late as $z = 20$ is required. In particular, the Universe is still transparent.

Physical mechanisms for reionization are considered by various authors (see *e.g.*, Tegmark, Silk and Blanchard 1993). CDM models are incapable of reionizing the universe earlier than $z \sim 100$. More generally, the recent constraints on the Compton $y$ parameter by FIRAS are beginning to constrain the epoch of reionization from being much earlier in high baryon density models. Our principal result in this Letter has been to demonstrate that over the entire allowed range of baryon density, the CMB spectrum on scales smaller than a few degrees is very sensitive to the thermal history of the universe.

## ACKNOWLEDGEMENTS

The authors would like to thank S. Dodelson for valuable discussions on the treatment of SP91 data and W. Hu for his help with calculations of the second order terms. This research has been supported at Berkeley in part by grants from NASA and NSF. N.S. acknowledges financial support from a JSPS Postdoctoral Fellowship for Research Abroad.

# Figure Captions

Fig.1 Power spectrum of temperature anisotropies for CDM models. Dashed line and solid line represent models with reionization at $z = 10$ and $z = 100$, respectively. The normalization is arbitrary.

Fig.2 Time evolution of temperature anisotropies, CDM and Baryon density fluctuations as a function of red-shift. The normalization is arbitrary. The scale is $k = 0.02 \mathrm{Mpc}^{-1}$ which corresponds to the Doppler peak. The ionization fraction is also shown.

Fig.3 Power spectrum of temperature anisotropies $\ell(2\ell+1)C_\ell/4\pi$ ((a)) as a function of $\ell$ for various reionization epochs. Contribution of the Vishniac effect in the optically thick universe is shown on large $\ell$ (labeled as 2nd). We multiply this effect by factor 10 since it is so small. Window functions $W_\ell$ for COBE, SP91 and OVRO are shown in (b).

Fig.4 Constraints on CDM models from SP91 in the $\Omega_B - h$ plane. The 95%-credible-limit is shown. The epoch of reionization is chosen as either $z = 10$ or $z = 20$. Lines corresponding to $\Omega_B h^2 = 0.01$ and $\Omega_B h^2 = 0.03$ are also shown.